\documentstyle[prl,aps,twocolumn,epsf,floats,psfig]{revtex}

\draft

\begin{document}

\twocolumn[\hsize\textwidth\columnwidth\hsize\csname
@twocolumnfalse\endcsname

\title{X-Ray-Diffraction Study of Charge-Density-Waves and Oxygen-Ordering in
YBa$_2$Cu$_3$O$_{6+x}$ Superconductor}

\author{Zahirul Islam$^{1,}$\cite{ZI} \and S. K. Sinha$^1$ \and
D. Haskel$^1$ \and J. C. Lang$^1$ \and G. Srajer$^1$
\and D. R. Haeffner$^1$ \and B. W. Veal$^2$ \and H. A. Mook$^3$}
\address
{$^1$Advanced Photon Source, Argonne National Laboratory, Argonne, IL 60439}
\address
{$^2$Materials Science Division, Argonne National Laboratory, Argonne,
IL 60439}
\address
{$^3$Oak Ridge National Laboratory, Oak Ridge, TN 37831}

\date{\today}

\widetext

\maketitle

\begin{abstract}
We report a temperature-dependent increase below 300 K of diffuse superlattice
peaks corresponding to {\bf q}$_0$=$\left(\sim\frac{2}{5},0,0\right)$
in an under-doped
YBa$_2$Cu$_3$O$_{6+x}$ superconductor (x$\approx$0.63). These peaks reveal
strong {\bf c}-axis correlations involving the CuO$_2$ bilayers, show a
non-uniform increase below $\sim$220 K with a plateau for $\sim$100-160 K,
and appear to saturate in the superconducting phase. We interpret this
unconventional $T$-dependence of the ``oxygen-ordering'' peaks as
a manifestation of a charge density wave in the CuO$_2$ planes
coupled to the oxygen-vacancy ordering.
\end{abstract}


\vskip0.5pc]
\narrowtext

The occurrence of ``stripes'' in the cuprate family of superconductors
has aroused considerable interest in recent years due to their possible
connection to the so-called `pseudogap' phase. The evidence of such a
phase appearing above the superconducting transition temperature has been
found from optical and transport properties  of the under-doped
superconducting compounds (see Ref.\ \cite{TT99} and references therein).
There has been much speculation about the microscopic nature
of the pseudogap phase \cite{BB00,CMV99,SC00} and a general concensus
that a thorough knowledge of this phase may be the key to understand
the mechanism of high-temperature superconductivity. One possibility
is that this phase is due to ordering of a kind ({\it e.g.} ``charge
stripes'') that competes with superconductivity and has important
implications for understanding the overall phase diagrams of these cuprates.
The first experimental observation of stripes was
in the Nd-doped La$_2$CuO$_4$ family with
neutron diffraction, where both anti-phase antiferromagnetic spin
stripes and the corresponding ``charge stripes'' ( at twice the wavevector of
the magnetic stripes) were observed\cite{JMT96,MvZ98}.
It is important to note that the diffraction
from ``charge stripes'' (or charge density waves (CDW))
is primarily from the atomic displacements or lattice
distortions associated with them. In the case of the Nd-doped compounds the
stripes are accepted to be static, stabilized by the low-temperature
tetragonal structure (LTT phase) of this compound. In the case of the
under-doped YBa$_{2}$Cu$_{3}$O$_{6+x}$ (YBCO) compounds, Mook and
collaborators \cite{HAM98} reported the existence of
incommensurate spin excitations at wavevectors of ($\pm$0.1, 0, 0) from
the ($\frac12$, $\frac12$, 0) reciprocal-space point which they
associated with dynamic stripe fluctuations. (Excitations originating
from the points at (0, $\pm$0.1, 0) relative to the
($\frac12$, $\frac12$, 0) point were ascribed to the twinned domains).
No static spin or charge density waves were found with neutrons,
although Mook and Dogan \cite{HAM99a} reported anomalies in phonons of
wavevector ($\pm$0.2, 0, 0) or the expected wavevector for the corresponding
dynamic charge stripes. 

X-ray scattering integrates over the energies of the fluctuations
at the wavevectors looked at, including dynamically fluctuating stripes.
Accordingly, in this work, we have used high-energy synchrotron x-rays
to search for charge stripes in under-doped YBa$_2$Cu$_3$O$_{6.63}$.
In this Letter we show that charge fluctuations
associated with stripes in the CuO$_2$ planes with wavevector
{\bf q}$_0$ = ($\sim\frac25$, 0, 0) relative to reciprocal-lattice points do
appear to develop at low temperatures, while scattering due to the
expected charge fluctuations at twice the wavevector of the incommensurate
spin-fluctuations ({\it i.e.} at ($\pm$0.2, 0, 0))
is weak or non-existent. We find that these charge-fluctuations
use ``oxygen-ordering'' (described below) as a template, and the wavevector
of the charge-fluctuation processes or CDW forming on the CuO$_2$ planes
are strongly correlated.

As in other cuprates, superconductivity in YBCO also arises from the
charge-carriers (or `holes') in the CuO$_2$ planes which form bilayers.
Unlike other cuprates, however, YBCO contains CuO$_x$-chain planes
(referred to as the basal or $ab$-planes) with variable number
of oxygens\cite{JDJ90}. The hole-density in the CuO$_2$ bilayers
is controlled by tuning the oxygen stoichiometry\cite{hole} 
which changes the number of oxygen atoms, or vacancies, in the $ab$-planes.
In the under-doped regime the study of YBCO becomes rather complicated
since these oxygen vacancies form various superstructures
\cite{JDJ87,CC87,RB89,NHA99,SY92} with ordering temperatures typically above
300 K\cite{JDJ87,NHA99,SY92}. These superstructures correspond to complicated
patterns, with well-defined modulations, such as ($\sim\frac25$, 0, 0)
relative to the nearest Bragg point  for x$\approx$0.63, along the {\bf a} axis,
of oxygen-full and oxygen-vacant CuO$_x$ chains ($\parallel$ {\bf b}-axis).
With the exception of the superstructure for x=0.5, the oxygen-ordering
(or vacancy-ordering) is rather short ranged\cite{NHA99,FY2000,VPla92}.
Since the ordering evolves with temperature via oxygen-diffusion
processes\cite{SJR89}, it is well-accepted that this ordering is essentially
frozen-in below room temperature \cite{NHA99,BVflux},
and should play no significant roles in influencing lower-temperature
properties of these cuprates. However, we find that the oxygen-ordering
phenomenon also helps to promote CDW instability in the CuO$_2$ planes
which may be associated with the pseudogap phase.

X-ray scattering experiments were performed at the Advanced Photon Source,
Argonne National Laboratory. A set of well-annealed,
self-flux-grown\cite{BVflux} crystals, was characterized on the SRI CAT 1-ID
beamline using a Weissenberg camera with 65 keV x-rays. A rectangular,
twinned crystal ($\sim$1000$\times$300$\times$70 $\mu$m$^3$) which showed no
obvious signs of diffraction peaks from extraneous phases was chosen from the
set. The magnetization measurements of this sample revealed $T_c$ to be 60 K
with a transition-width of $\sim$0.5 K, suggesting a high-degree of
compositional homogeneity in the bulk of the sample. Most of the
present work was carried out on the SRI CAT 4-ID beamline using
36 keV x-rays. A Si-(1, 1, 1) reflection was used as a monochromator
with the undulator $5$th harmonic tuned to provide maximum flux at this
energy. The sample was cooled in a closed-cycle He refrigerator. The
temperature-dependent measurements were carried out on warming from low
temperature. Y-fluorescence was carefully monitored to normalize to
the same diffracting volume at every temperature.

\begin{figure}[h!t]\centering
\epsfxsize=75mm
\epsfbox{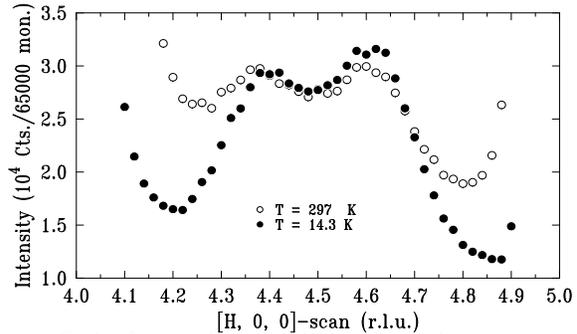}
\caption{Comparison of diffuse scattering at room temperature and 14.3 K,
respectively, showing a clear enhancement at low temperature.}
\label{df-14vs300}
\end{figure}

The primary direction of interest in the reciprocal space is ${\bf a}^*$
({\it i.e.} the direction along the shorter Cu-O-Cu links in
the CuO$_2$ planes and perpendicular to the CuO$_x$ chains \cite{JDJ90})
along which ``oxygen-ordering'' superlattice peaks appear and phonon
line-broadening peaks have been reported \cite{HAM99a}. Fig.\ \ref{df-14vs300}
shows reciprocal lattice scans ({\it i.e.} [H, 0, 0] scans) between
(4, 0, 0) and (5, 0, 0) Bragg peaks,
respectively, at two different temperatures. Two broad superlattice
peaks due to oxygen ordering characterized by a modulation vector of
{\bf q}$_0$ = ($\sim\frac25$, 0, 0) are clearly observed at both
temperatures. What is intriguing is that at the lower
temperature (14.3 K) these peaks are strongly enhanced whereas
the thermal diffuse scatterng (TDS) is significantly reduced. This enhancement
is clearly at odds with conventional expectations and can not be
accounted for by oxygen ordering alone. We note that there
appears to be {\it no} significant scattering above the diffuse tails
near the expected incommensurate charge-fluctuation peaks at
($\sim$4.2, 0, 0) and ($\sim$4.8, 0, 0), respectively\cite{HAM99a}. 
Furthermore, at the lower temperature, a series of systematic scans
along {\bf a}$^*$ with different values of K
revealed diffuse peaks corresponding to {\bf q}$_0$, with no signs
of peaks related to ($\sim$0.2, 0, 0). In addition, considering the
possibility of staggering of the charge fluctuations between the CuO$_2$
planes within a bilayer, we have collected scans along {\bf a}$^*$,
with L values ranging from -2 to 0 in increments of 0.2. No clear
signs of peaks at H$\approx$4.2 or at H$\approx$4.8, respectively,
were found in these scans either.

The intensities (corrected for background) of the {\bf q}$_0$-peaks are
of the order of $\sim$500-2500 counts/sec. at 14.3 K, some $\sim10^6-10^7$
orders-of-magnitude weaker than those of the Bragg peaks. They are due to
local-lattice-distortions\cite{TZ92,VPla92} which at high
temperatures have been associated with oxygen-vacancy
ordering as the lattice relaxes around the vacancies. The contribution of a
lattice displacement $\delta {\bf u}$ to the diffuse scattering amplitude
at a momentum transfer {\bf Q} varies as {\bf Q}$\cdot \delta${\bf u}.
A series of [H, 0, L] scans with H ranging from
0.1 to 0.9 collected for a given L value of 6, 7, or 8, respectively, did not
show any significant scattering around ($\sim\frac25$, 0, L), implying that
$\delta{\bf u}\perp{\bf c}$. Further, the absence of any significant scattering
around (0, 8, 0)$+${\bf q}$_0$ suggested $\delta{\bf u}\perp{\bf b}$.
Indeed, a series of [H, 0, 0] scans revealed that
$\frac{I({\bf\text{G}+\bf\text{q}}_0)}{I({\bf\text{G}-\bf\text{q}}_0)}\propto
\frac{|\bf\text{G}+\bf\text{q}_0|^2}{|\bf\text{G}-\bf\text{q}_0|^2}$,
where $I$ is the intensity at a
{\bf Q}={\bf G}$\pm${\bf q}$_0$ and {\bf G} is a reciprocal
lattice vector, confirming that the atomic displacements
are longitudinal ($\parallel{\bf a}$).

\begin{figure}[h!t]\centering
\epsfxsize=75mm
\epsfbox{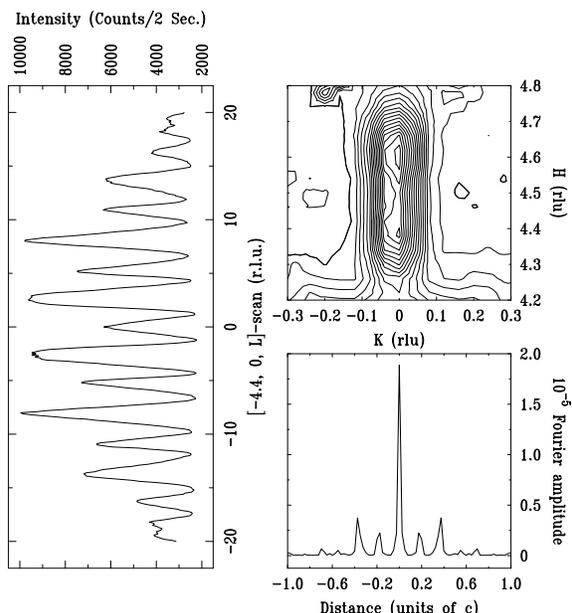}
\caption{Top right: Two dimensional scan showing the extent of diffuse
scattering in the [H, K, 0] zone. Left: The modulation of [-4.4, 0, 0]
diffuse peak along {\bf c*}. Lower right: The Fourier transform showing
distances between planes that are correlated. All the data were collected
at 14.3 K.}
\label{hkmesh-fft}
\end{figure}

Fig.\ \ref{hkmesh-fft} (top-right panel) shows a two dimensional mesh
revealing the extent of the diffuse scattering in the [H, K, 0] zone. The
profile is elongated along {\bf a*} with two concentrated
summits at (4.4, 0, 0) and (4.6, 0, 0), respectively. The width of the
peaks along {\bf b*} is much sharper ($\sim0.05$ r.l.u.) than that along
{\bf a*} ($\sim$0.20 r.l.u.) suggesting a slightly longer-range
correlation ($\sim$30 {\AA}) along the CuO$_x$ chains than the
correlation ($\sim$7 {\AA}) along {\bf a}. Such an anisotropy of
the correlation lengths within the {\it ab}-plane is consistent with
quasi-one-dimensional nature of the underlying distortions. We note that
we have eliminated the ambiguity between ($\frac25$, 0, 0) and
(0, $\frac25$, 0), respectively, due to twinning (see also
\cite{RB89,VPla92}). All the diffuse peaks are equidistant from
the nearest (H, 0, 0) Bragg peak, and the splitting of the
(H, 0, 0) peaks and the corresponding (0, H, 0) peaks from
the twin were well-resolved in our current set-up.

The left panel in Fig.\ \ref{hkmesh-fft} shows how the diffuse
scattering at (-4.4, 0, 0) is modulated along {\bf c*}, the
significance of which was not fully appreciated \cite{VPla92,PS95}
until now. This intensity-modulation is not purely sinusoidal
suggesting that the displacements on more than two scattering planes
are correlated along the {\bf c}-axis. If we describe the scattering
due to atomic displacements ($\parallel{\bf a}$)
in a given layer $n$ ($\perp{\bf c}$) by a complex structure
factor $F_n$, then the intensity modulation along {\bf c*}
of the diffuse peak at {\bf Q}=($Q_x, 0, Q_z$) is given by
\begin{eqnarray*}
I(Q_z)=CQ^2_x\left(
\sum_{n,n'}<F_nF^*_{n'}>e^{-iQ_z(z_n-z_{n'})}\right)
\end{eqnarray*}
where $z_n$ is the height along {\bf c}-axis of layer $n$. So, the heights
of the correlated planes can be obtained by a direct Fourier transform
of the intensity modulation. As shown in Fig.\ \ref{hkmesh-fft}
(bottom panel) there are two planar distances, $z_1$ = 0.362$\pm$0.008 $c$
and $z_2$ = 0.187$\pm$0.008 $c$, respectively, that are correlated.
$z_1$ corresponds to the distance between the CuO$_2$
and CuO$_x$-chain planes, whereas $z_2$ is consistent with both the
CuO$_2$-O$_{\text{apical}}$ (O$_{\text{apical}}$ is the oxygen atom
in between Cu atoms along the {\bf c} axis) and the chain-Ba distances,
respectively\cite{JDJ90}. Note that the correlation between the
CuO$_2$-plane and CuO$_x$-chain ($z_1$) is clearly much stronger.

\begin{figure}[h!t]\centering
\epsfxsize=70mm
\epsfbox{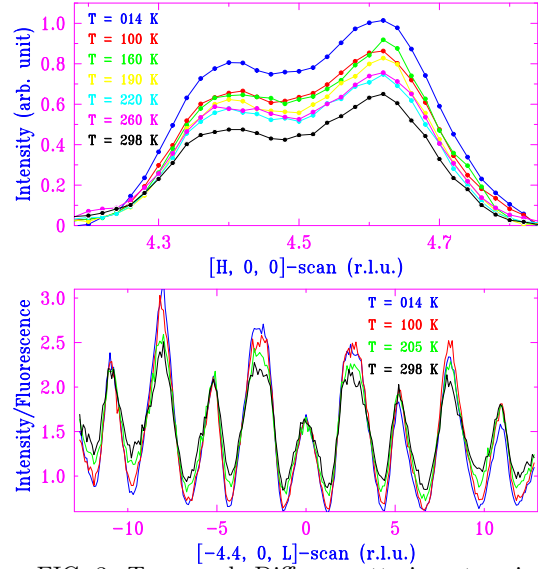}
\caption{Top panel: Diffuse scattering at various temperature. The background
including the TDS has been removed from the data. Bottom panel: L-modulations
of (-4.4, 0, 0) diffuse peak at selected temperatures. The intensity
modulations are clearly enhanced as $T$ is lowered.}
\label{df-dat}
\end{figure}

The measurements of the diffuse peaks and their L-modulations
at various temperatures are shown in Fig.\ \ref{df-dat}.
The background, including the TDS, was modeled using two Lorentzian tails
and a linear term, respectively,  and was subsequently removed from
the data shown in Fig.\ \ref{df-dat} (top panel). The diffuse
scattering and the modulation (Fig.\ \ref{df-dat}: bottom panel)
both show clear indication of increase at low $T$. Interestingly,
Fig\ \ref{df-dat} (top panel) also indicates the presence of two
temperature regions, $\sim$100-160 K and $\sim$220-260 K, respectively, where
there are no significant differences in the integrated intensity (proportional
to the area under the curve) with temperature.

\begin{figure}[h!t]\centering
\epsfxsize=70mm
\epsfbox{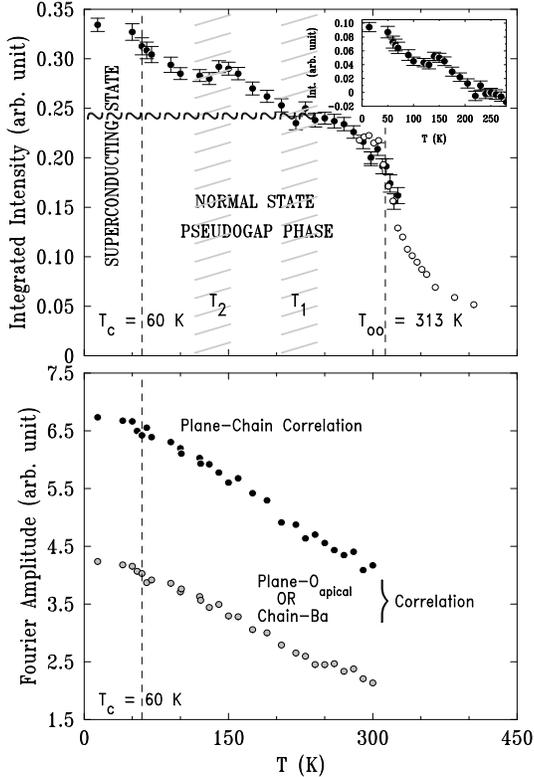}
\caption{Top: $T$-dependence of the diffuse scattering at
(4.4, 0, 0). ($\circ$) are data from Andersen and co-workers
(see text) normalized to overlap
with our data for 300-328 K. ($\sim$) depict low-$T$ extrapolation of the
intensity between $\sim$220-260 K. Hatched areas identify $T$-regions
where breaks in the intensity are observed. Inset: ``Order parameter''
of the charge stripes. Bottom: $T$-dependence of the correlations.}
\label{df-Tdep}
\end{figure}

Fig.\ \ref{df-Tdep} summarizes the $T$-dependence of the diffuse
scattering. We have included the data above 300 K from
Ref.\ \cite{NHA99} to illustrate the expected behavior in the
higher-$T$ region, inaccessible in our measurements.
Since the data from Ref.\ \cite{NHA99} were for oxygen stoichiometry
of 6.67 with an ordering $T$ of 323 K we have shifted
the temperature down by
10 K to be consistent with the transition temperature, $T_{oo}$ $\sim$ 313 K,
for our composition. As shown, the intensity gradually increases
on lowering $T$ and becomes nearly constant in the $\sim$220-260 K range.
Below $T_1\sim$220 K, however, the intensity starts to rise beyond the
extrapolation (tildes) of the 220-260 K intensity, until $T_2\sim$160 K
below which it seems to level off. Then, as  the superconducting
state is approached the intensity gradually starts to rise again
and appears to saturate in that phase. The inset shows the intensity
after subtracting off the 220-260 K value. The bottom panel shows
that the Fourier amplitudes of $z_1$ and $z_2$ grow
at low $T$ and saturate in the superconducting state. Within our
experimental uncertainties, we did not observe any changes in
$z_1$ or $z_2$.

In summary, our experiments showed that the diffuse peaks corresponding
to ${\bf q}_0$=($\sim\frac25$, 0, 0) originate from mutually-coupled,
longitudinal local-lattice-displacements of the CuO$_2$, CuO$_x$-chain,
and Ba or/and O$_{\text{apical}}$, respectively.
The primary result of our study is the
unconventional $T$-dependence of this diffuse scattering much below
the room temperature where the oxygen-vacancy ordering is frozen-in.
We note that the anomalous low-temperature increase of the scattering
is below $T_1$, which roughly corresponds to the entry into the
``pseudogap'' phase in this material. In addition, there appears to
be a plateau at $T_2$. Recent ion-chanelling measurements
observed similar characteristic temperatures in the $T$-dependence
of incoherent atomic displacements in under-doped YBCO\cite{RPS00}.
It is interesting to note that $\mu$SR and neutron diffraction studies
\cite{JES01,HAM01} on under-doped YBCO suggested the onset of
$d$-density wave order \cite{SC01} below $T_2$.

In conclusion we find that in carefully prepared samples
of YBa$_2$Cu$_3$O$_{6.63}$, the CuO$_x$-chain vacancies show significant
short-range order as rows of missing oxygen atoms and become
strongly correlated with instabilities in the CuO$_2$ planes
that grow into the pseudogap phase. In our view, stripe-like
patterns of displacements form in the CuO$_2$ planes at low $T$ as the CDW
evolves using the underlying oxygen-ordering as a template. Fig.\ \ref{df-Tdep}
shows how the ``order parameter'' of such a stripe phase may evolve as the
pseudogap phase and the superconducting state, respectively, are entered.
In this connection we note that recent neutron diffraction studies
on a single crystal YBCO sample \cite{MA00} show a peak
at ($\sim$0.2, 0, 0). The absence of this peak in the
x-ray scattering leads us to speculate that this peak is in fact
due to the magnetic stripe corresponding to our observed scattering at
twice this wavevector, and also commensurate with the oxygen ordering in
the chains. Other incommensurate charge fluctuations such as those
associated with the spin fluctuations around ($\sim$0.1, 0, 0)
if they exist may be too rapid for the lattice distortions to follow
and consequently we would not have seen them. A further consequence
would be that in poorly prepared samples where the oxygen chain
vacancies were disordered, the CuO$_2$
stripes would not exist, or be themselves disordered. It would be
interesting to look for stripes in other cuprates where oxygen
vacancy ordering is absent.

Use of the Advanced Photon Source is supported by the U.S. Department
of Energy, Office of Science, Office of Basic Energy Sciences, under
Contract No. W-31-109-ENG-38. We thank J. Almer for his help with the
Weissenberg camera. We have benefitted from discussions with
A. Paulikas (MSD) at Argonne.

\end{document}